\documentstyle[amsfonts,epsfig,aps,prd]{revtex}

\newcommand\NPB{{Nucl. Phys.} B}

\newcommand\PLB{{Phys. Lett.} B}

\newcommand\PRL{Phys. Rev. Lett.}
\newcommand\PRC{{Phys. Rev.} C}
\newcommand\PRD{{Phys. Rev.} D}

\newcommand\ZPC{{Z. Phys.} C}

\newfam\BMath
\font\BMathL=cmmib10 
\font\BMathl=cmmib7
\font\BMathm=cmmib5
\textfont\BMath=\BMathL \scriptfont\BMath=\BMathl
\scriptscriptfont\BMath=\BMathm

\renewcommand\P{{\fam\BMath p}}

\newcommand\K{{\fam\BMath k}}

\newcommand\g{\gamma}
\renewcommand\d{\delta}

\newcommand\q{\theta}

\newcommand\m{\mu}
\newcommand\n{\nu}

\newcommand\p{\pi}

\newcommand\f{\phi}

\renewcommand\j{\psi}

\newcommand\cf{{\cal F}}

\newcommand\cl{{\cal L}}
\newcommand\cm{{\cal M}}   
\newcommand\cn{{\cal N}}
\newcommand\co{{\cal O}}   
\newcommand\cp{{\cal P}}   

\newcommand\cs{{\cal S}}

\renewcommand\exp{\mbox{\rm exp}}  

\newcommand\lra{\longrightarrow}

\newcommand\del{\partial}

\newcommand{\half}{\frac{1}{2}}

\newcommand\be{\begin{equation}}
\newcommand\ee{\end{equation}}
\newcommand\bea{\begin{eqnarray}}
\newcommand\eea{\end{eqnarray}}

\newcommand\ba{\begin{array}}
\newcommand\ea{\end{array}}

\newcommand\eref[1]{Eq.~(\ref{#1})}
\newcommand\etwref[2]{Eqs.~(\ref{#1}) and (\ref{#2})}

\newcommand\eforef[4]{Eqs.~(\ref{#1}), (\ref{#2}), (\ref{#3}) and (\ref{#4})}

\newcommand\bfi{\begin{figure}}
\newcommand\efi{\end{figure}}
\newcommand\bpi[1]{\begin{picture}#1}
\newcommand\epi{\end{picture}}

\newcommand{\fref}[1]{Fig.~\ref{#1}}
\newcommand{\ftwref}[2]{Fig.~\ref{#1} and \ref{#2}}

\def\jou#1#2#3#4{{#1} {\bf #2}, #3 (#4)}

\font\euf=eufm10
\font\eufs=eufm7
\def\es{\mbox{\euf s}}
\def\ess{\mbox{\eufs s}}
\def\eF{\mbox{\euf F}}

\def\fm{f^{(-)}}
\def\fp{f^{(+)}}
\def\tfm{\tilde f^{(-)}}
\def\tfp{\tilde f^{(+)}}

\begin{document}


\draft
\wideabs{

\title{Disentangling the Imaginary-Time Formalism at Finite Temperature} 

\author{S.M.H. Wong}

\address{
School of Physics and Astronomy, University of Minnesota, Minneapolis, 
Minnesota 55455
}

\maketitle

\begin{abstract} 
We rewrite the imaginary-time formalism of finite temperature 
field theory in a form that all graphs used in calculating 
physical processes do not have any loops. Any production of a particle 
from a heat bath which is itself not thermalized or the decay and 
absorption of a similar particle in the bath is expressed entirely in 
terms of the sum of particle interaction processes. These are themselves 
very general in meaning. They can be straight forward interactions or 
the more subtle and less well-known purely interference processes 
that do not have a counterpart in the vacuum. 
\end{abstract}

\pacs{PACS: 11.10.Wx, 12.38.Bx, 12.38.Mh \hfill NUC-MINN-00/08-T}
}

\section{Introduction}
\label{s:intro}

For equilibrium field theory at finite temperature, there are two main 
methods of performing calculations. They are the imaginary-time formalism 
where one starts out in Euclidean space and analytically continues back
into Minkowski space at the end of the calculation, and the real-time
formalism where the calculation is done in Minkowski space with real 
time throughout. Because the latter has explicit real time dependence, 
it is therefore more suitable for time-dependent problems. However 
it also has the feature of the doubling of the field degrees of 
freedom so that each field acquired a partner and the propagators 
became 2x2 matrices, therefore in the sense that the components used
in calculations are scalar quantities\footnote{Here we are considering
everything in the zero temperature field theory as scalar quantities,
in other words for the purpose of discussion here, we make no distinctions 
between scalar, spinor, and vector fields related quantities. These
are all scalar in the sense they have no doubling of the degrees of
freedom in the vacuum.} at $T=0$ versus matrix quantities in the 
real-time formalism the simple analogy and straightforward similarity 
to field theoretical calculation in the vacuum are absent. Intuitively, 
the difference between the calculation at zero and finite temperature 
should only be that the latter acquires thermal weights in the phase 
space integrations. This clearly is not the case in the real-time 
formalism because one has in addition to deal with matrix quantities. 
In this regard, the imaginary-time formalism resembles much more 
the zero temperature field theory. Having said that it must 
be stated that we are well aware of the Braaten-Pisarski 
resummation where perturbation theory at finite temperature must be 
rearranged \cite{rp,bp1,bp2,ft,tw} so that it is not just a matter of 
thermal weights between zero and finite temperature. There are various
applications of this resummation scheme, see for example the above
references and also \cite{bt,bnnr,w1,w2,thoma}. These have also been 
recast into the form of kinetic equations for soft particles 
within a heat bath \cite{bi1,bi2,ja1,keetal}. We will choose not to 
complicate matters in this paper and leave resummation out, not even 
including it partially as in \cite{wel0,dh,aa1,aa2}, or in other words we 
will consider only particles typically at the same scale as the temperature 
or higher when resummation is formally not necessary. We will address  
the imaginary-time formalism itself and deal with a number of issues 
on its usage. Because of its intrinsic similarity with the vacuum 
theory, it should be possible to bring it into the form where the 
difference is essentially in the thermal weights. It will be shown,
however, that there is a limit to how far this resemblance will go.
And we will go even further by cutting all internal loops. 

The imaginary-time formalism is a very nice formalism in that by 
following the established calculational rules of replacing the thermal
discrete imaginary energy sum by contour integrals and analytic 
continuation, the calculations can be done very similarly to the familiar
vacuum field theory \cite{jbook}. The needed thermal distributions will be 
there at the end and repeated self-energy insertion along a single line will 
be sorted out even if that line is subsequently cut or put on-shell
provided the rules are followed correctly. Thus the formalism is very compact
which can hide many physical processes. For this reason, although the
formalism is quite superior mathematically, it can be very unclear 
when it comes to finding out exactly what physical processes are involved 
in a particular calculation. The mathematical advantage becomes the 
disadvantage when it comes to the physics. For each given calculation 
involving medium modification of the properties of a particle or the 
production of one, it should be possible to write the result in a form that
is a sum of all the individual contributing processes. These should 
apparently be the phase space integral of each relevant process weighed 
by a product of thermal distributions but this is not entirely true as 
has already been mentioned and will be clear later.

Another related aspect in thermal field theory that has not been 
categorically pointed out and discussed is the much richer possibility 
of interference when a process happens inside a heat bath. Although 
interference graphs are well-known to be necessary both at zero 
and non-zero temperature, they have largely kept their $T=0$ form 
when being discussed within the finite $T$ context \cite{ks2,wel2}. Thus 
much of the richness was not obvious and remained hidden within the 
``simple'' zero temperature Feynman graphs. The authors are only aware 
of infrequent mentions of this here and there, see ref. \cite{lt,ls} 
which are two of the few papers which, as far as the aspects of
the possibility of interference is concerned, had gone to some depths. 
This may be the case because the existence of spectator particles easily 
masks any forward scattering or similar processes and so ensuring that 
it is impossible to tell whether the latter happened or not. 
When we say forward scattering here, we mean it in a generalized 
sense because this can happen to both a fermion or a boson. We will 
discuss this below, pursue further in the direction of \cite{lt,ls} 
but within the imaginary-time formalism and use thermal QED coupled to 
a massive vector particle which is itself not part of the heat bath as 
an example in Sect. \ref{s:ceg}.

\section{Thermal process = Vacuum Process $\otimes$ Thermal Weights, \\
Almost But Not Quite}
\label{s:teqvw}

In thermal calculations, it very often concerns the rate of production 
of some particles from the heat bath which are themselves not thermalized 
or how the medium modifies the properties of a particle such as its 
decay or absorption inside the hot medium. When this is done within the 
imaginary-time formalism, one can calculate this order by order 
(again we are ignoring resummation here) and the formalism will yield
formulae involving thermal distributions and other expressions. 
This is fine if one is interested only in the answer. If on the other
hand, one would like to know what physical processes contribute to the
production of a particle or how another get absorbed or stimulated to 
decay inside a medium, one can express the formalism in another more 
physically explicit way but equivalent to the original formalism. 
To this end, we now makes the claim that the differential production or 
emission rate of a particle with 4-momentum $k$ per unit volume per unit 
time from a heat bath, but which is itself not in the heat bath, is 
essentially given by the discontinuity of the self-energy as 
\be 2 k^0 \frac{d R}{d^3 k} =
    -{{i\;\mbox{Disc}\; \Pi (k)} \over {\exp(k^0/T)-1}} = I(k) +J(k)
\label{eq:pb}
\ee
for a scalar boson (the generalization to vector bosons is straight 
forward) and
\be 2 k^0 \frac{d R}{d^3 k} = 
    -{{i\;\mbox{Disc}\; \Sigma (k)} \over {\exp(k^0/T)+1}} = I(k) +J(k)
\label{eq:pf}
\ee
for a fermion. Each of these can be explicitly expressed in terms of a sum of 
all the physical processes contributing to the production of this particle
represented by the function $I(k)$ and $J(k)$ on the right-hand-side (r.h.s.). 
The $I(k)$ part, because of its resemblance to what one gets by cutting 
rules at zero temperature \cite{cut,elop}, is known (see below). 
It is essentially the sum of the relevant phase space integrals over the 
allowed kinematical range of the probability for each production process 
weighed by the particle distributions. The existence of the $J(k)$ 
part is, however, not well-known at least not to the same extent
that it will be expressed later on in this paper. It is very easy 
therefore to assume erroneously that there is only the contribution 
from $I(k)$. This is the case for example in the papers \cite{ks2,ks1} 
We will clarify this in the following sections. 

For those processes that do not involve spectator interference graph 
(see Sect. \ref{s:is} for clarification) and therefore whose probabilities 
can be written as squared modulus of the corresponding matrix elements, 
they are represented by  
\bea I(k) 
   &=& \sum_\cp \int d \Phi_\cp \; (2\p)^4 \d^4 (-k+\sum_{i\in \cp} s_i\, p_i) 
                                                 \nonumber \\
   & & \;\; \times |\cm_\cp|^2 \; F_\cp          
\label{eq:i}
\eea 
Here in the energy-momentum conserving delta function for each process 
$\cp$, there are the signs $s_i$ which depend on whether each 4-momentum $p_i$ 
is incoming $s_i=+$ or outgoing $s_i=-$. $F_\cp$ is a product of particle 
distributions for each of the participants that has {\em an entry} in the 
energy-momentum conserving delta function 
\be F_\cp = \prod_{i \in \cp} \; s_i\; \big (\q(s_i)+\es_i\; \q(-s_i) \big ) 
    \; f^{(\ess_i)}(p^0_i)   \; , 
\label{eq:distr}
\ee
where we used $f^{(+)}$ and $f^{(-)}$ to denote Bose-Einstein and Fermi-Dirac 
distribution respectively and $\es_i$ is a sign for this purpose. 
The measure $d \Phi_\cp$ represents that of the phase space integrations 
of the process $\cp$  
\be d \Phi_\cp = \prod_{i \in \cp} \frac{d^4 p_i}{(2\p)^4} \;
                 \d^{(s_i)} (p_i^2 -m_i^2)   \; .
\label{eq:psm}
\ee
We have used this $I(k)$ above and will use it again below for generic 
representation of this kind of sum of the contributing processes of 
the squared modulus type. 

The other generic function $J(k)$ is for the sum of interference 
processes that involve spectator particles in either one probability 
amplitude of the convoluting pair and the emission-absorption or vice 
versa of the same particles in the other amplitude of the pair. The 
function can be represented by  
\bea J(k) 
   &=& \sum_{\cp'} \int d \Phi'_{\cp'} \; d \mbox{\boldmath $\varphi$}_{\cp'}
       (2\p)^4 \d^4 (-k+\sum_{i\in \cp'} s_i\, p_i)          \nonumber \\
   & & \;\; \times (\bar \cm^{}_{\cp'} \cm^*_{\cp'} +\bar \cm^*_{\cp'} \cm^{}_{\cp'}) 
       \; F_{\cp'} \; \cf_{\cp'} \; .
\label{eq:j}
\eea
Because in these interference graphs, there are emission-absorption of 
particles of the same momentum in a single amplitude and some of these 
possibilities have their origin in thermal self-energy insertions on the 
external lines, which can also be viewed as generalized forward scattering
on the external lines, the phase space integration measure 
$d \Phi'_{\cp'}$ for such a typical process $\cp'$ in general becomes 
\bea d \Phi'_{\cp'} &=& \prod_{i \in \cp', i\not \in \cs} \frac{d^4 p_i}{(2\p)^4} \;
                        \d^{(s_i)} (p_i^2 -m_i^2) \;        \nonumber \\
     & & \times  \prod_{j \in \cs} \frac{d^4 p_j}{(2\p)^4} \; 
                 (-1)^{n_j} \d^{(n_j) (s_j)} (p_j^2 -m_j^2) 
\label{eq:phint}
\eea 
for a subgroup $\cs$ of all the external lines of the process $\cp'$ in
which each member $j$ in the subgroup has $n_j$ number of thermal 
self-energy insertions. We find it clearer here to say thermal 
self-energy insertion since this should be familiar to the readers, 
but we will eventually switch to the new meaning of the occurrence of 
any generalized forward scattering on the external line $j$. 
This will be further explained below in Sect. \ref{s:is}. 
The above phase space integrations only take care of those momenta that
enter into the overall energy-momentum conserving delta function. There
are other momentum integrations that usually being labeled as loop-momenta. 
These are represented by the other measure $d \mbox{\boldmath $\varphi$}_{\cp'}$
in \eref{eq:j} above. Here it suffices to state that we do not consider 
loops as such because each loop can be opened up and interpreted as 
emission-absorption of particles with the same 4-momentum. This 
interpretation will allow us to treat these loop-momentum integrals as 
phase space integrals so the measure $d \mbox{\boldmath $\varphi$}_{\cp'}$ 
can be expressed in general in a form very similar to \eref{eq:phint}. 
These other phase space integrations will therefore acquire thermal 
distributions as well. The product of these is $\cf_{\cp'}$ in 
\eref{eq:j}. This is different from $F_{\cp'}$ which retains the form 
in \eref{eq:distr} for the external lines.  

For the decay or absorption of a particle with momentum $k$ and 
mass $M$ not thermalized in the medium, we can similarly write 
\be 2 M \; \Gamma = 
    \;+{{i\;\mbox{Disc}\; \Pi (k)} \over {\exp(-k^0/T)-1}} = I(-k) +J(-k)
\label{eq:db}
\ee 
for a boson and 
\be 2 M \; \Gamma = 
    \;-{{i\;\mbox{Disc}\; \Sigma (k)} \over {\exp(-k^0/T)+1}} = I(-k) +J(-k)
\label{eq:df}
\ee
for a fermion with the same mass. We are treating the absorption as
stimulated decay in the presence of the medium, hence the width here
$\Gamma$ is the width in the medium. The thermal factor in the 
denominators of \eforef{eq:pb}{eq:pf}{eq:db}{eq:df} are correct
because in the limit $T\lra 0$, there is no longer a medium to
produce any particle so $dR/d^3 k \lra 0$. The width, however, will
remain finite and is given now by the discontinuity of the self-energy
in the vacuum.

\section{Simple Examples}
\label{s:seg}

The relation of the discontinuity of the self-energy to the sum of 
phase space integrals over each contributing process given in the 
previous section can be readily shown to hold for simple cases. We 
review the case of the one-loop self-energy as a simple example. For 
the production of a massive fermion from other massless fermions and bosons
in a heat bath, the required loop graph is the one of the usual fermion 
self-energy. Within the imaginary-time formalism, we write it as 
\be \Sigma(k) = T \sum_{p_4 = i p_0} \int \frac{d^3 p}{(2\p)^3} 
                \frac{\cn (k,p)}{p^2 \; (k-p)^2}  \; .
\ee
Here we have not stated explicitly which theory is being considered
except by using the usual fermion self-energy graph, it has to be one with 
vector coupling between boson and fermion-antifermion. The numerator 
is simply denoted by a function $\cn$ since its details are not required
here. After converting the discrete energy sum into contour integration
and performing the latter \cite{jbook}, one gets 
\bea \Sigma(k) =&& -\int \frac{d^3 p}{(2\p)^3}                      \nonumber \\
    \times 
    \bigg \{ && \frac{(\half+\fp(p))}{2p} \frac{\cn(k,p)}{(k-p)^2} \Big |_{p_0 =p}
                                                                    \nonumber \\
            +&& \frac{(\half+\fp(p))}{2p} \frac{\cn(k,p)}{(k-p)^2} \Big |_{p_0 =-p}
                                                                    \nonumber \\
            +&& \frac{(\half-\fm(|\K-\P|))}{2|\K-\P|} \frac{\cn(k,p)}{p^2} 
                \Big |_{p_0 = k_0+|\K-\P|}                          \nonumber \\
            +&& \frac{(\half-\fm(|\K-\P|))}{2|\K-\P|} \frac{\cn(k,p)}{p^2} 
                \Big |_{p_0 = k_0-|\K-\P|}                          
     \bigg \}      \; .
\label{eq:fse}
\eea
The vacuum part has also been included in the above expression. 
Now performing analytic continuation and taking the discontinuity 
of the self-energy $\Sigma(k)$ with the delta function $\d(k_0+p+|\K-\P|)$ 
for the production of the massive fermion in mind, only the second
and the third term in \eref{eq:fse} have the right discontinuity. 
For production, we have to set $k^0\lra -k^0$ to get 
\bea i \mbox{Disc} \Sigma(k) &&= -\int \frac{d^3 p}{(2\p)^3} 
       (2\p) \d (-k_0+p+|\K-\P|)                                    \nonumber \\ 
 \times && \frac{\big (1+\fp(p)-\fm(|\K-\P|) \big )}{2p\;2|\K-\P|} 
        \cn (-k,p) \Big |_{p_0 =p}  \; .                            \nonumber \\ 
\eea
This can be rearranged to
\bea && -\frac{i\mbox{Disc} \Sigma(k)}{\exp(k_0/T)+1}               \nonumber \\
       =\int && \frac{d^4 p}{(2\p')^4}  (2\p) \d^{(+)} (p^2)      
                \frac{d^4 p'}{(2\p')^4} (2\p) \d^{(+)} (p'^2)       \nonumber \\ 
      \times && (2\p)^4 \d^{(4)} (-k+p+p')                          \nonumber \\ 
      \times && \fp(p_0) \fm(p_0') \; \cn (-k,p)                    \nonumber \\
       = \;\;\;\; && 2 k_0 \frac{d R}{d^3 k} 
         \Big |_{\mbox{\scriptsize one-loop}} \; ,
\eea
the form we expected from \eref{eq:pf}. The $\cn$ is essentially just 
the squared of the vector-fermion-antifermion coupling, which is 
all there is for the squared modulus of the probability amplitude at 
leading order. 

One could easily use \eref{eq:fse} to obtain the discontinuity for
the decay of a massive fermion into a fermion and a boson in the heat
bath. The energy conserving delta function to get from \eref{eq:fse}
in this case is $\d (k_0-p-|\K-\P|)$. Only the first and last term
from \eref{eq:fse} contribute. It is simple to follow the same steps
and arrive at the form given in \eref{eq:df} or
\bea && -\frac{i\mbox{Disc} \Sigma(k)}{\exp(-k_0/T)+1}              \nonumber \\
       =\int && \frac{d^4 p}{(2\p')^4}  (2\p) \d^{(+)} (p^2)      
                \frac{d^4 p'}{(2\p')^4} (2\p) \d^{(+)} (p'^2)       \nonumber \\ 
      \times && (2\p)^4 \d^{(4)} (k-p-p')                           \nonumber \\ 
      \times && \big (1+\fp(p_0) \big )\big (1-\fm(p_0') \big) 
                \; \cn (k,p)                                        \nonumber \\
       = \;\;\;\; && 2 M \; \Gamma \big |_{\mbox{\scriptsize one-loop}} \; .
\label{eq:gam_ol}
\eea
Note that this could equally have been written with $\d^{(-)}$ with $f$
as given in the previous section instead of $\d^{(+)}$ with $1\pm f$. 
Both of the above explicit examples have not the $J(k)$ function part
because this will come at higher orders. More precisely these 
contributions come in at two-loop and higher orders.

\section{General Proof} 
\label{s:gp}

In this section in order to put up a framework so that later sections
can be more readily understood, we show in our own way that 
\eforef{eq:pb}{eq:pf}{eq:db}{eq:df} hold in general. We saw in 
Sec. \ref{s:teqvw} that both the $I(k)$ and $J(k)$ functions 
possess the same structure of $I(k)$ at least for the external 
lines included in the delta function conserving the overall
4-momentum. That is as far as the part associated with lines of the 
contributing graphs that carry part of the total energy-momentum, the
two functions are the same. $J(k)$ has some extra phase space integrals 
and distributions but these have little to do with the overall momentum
conservation and they can be viewed as substructures. This will be 
discussed later on. First the main structure of 
\eforef{eq:pb}{eq:pf}{eq:db}{eq:df}, which is the structure of the 
function $I(k)$, will be shown first. 

For a self-energy at arbitrary order in the coupling of the theory, 
it can always be rearranged into the form shown in \fref{f:se} where
momentum $p_1$ flows through line 1, $p_2$ flows through line 2 ... 
etc and the last line $m$ carries the boson or fermion momentum 
$k$ of the self-energy under consideration and the sum of the other 
lines above it so that $p_m = k+\sum_i^{m-1} p_i$. The two blobs 
shaded differently need not be the same in general and they can 
be any graphs from very simple to very complex connecting the lines 1 
to $m$. Our aim is to put all these lines on-shell so that they become 
the external lines and each has an entry in the overall 4-momentum 
conserving delta function. In other words, we are dividing the 
self-energy in two by cutting through line 1 to $m$. There are of 
course more than one way to group lines into the form in \fref{f:se} 
and therefore different cuts possible on that graph. Also at any
given order, the self-energy will be a sum of graphs of the form of
\fref{f:se}. The full results must therefore be a sum both over different
graphs and cuts on those graphs. This sum will eventually become our 
sum over different contributing processes. 

\bfi
\centerline{
\epsfig{figure=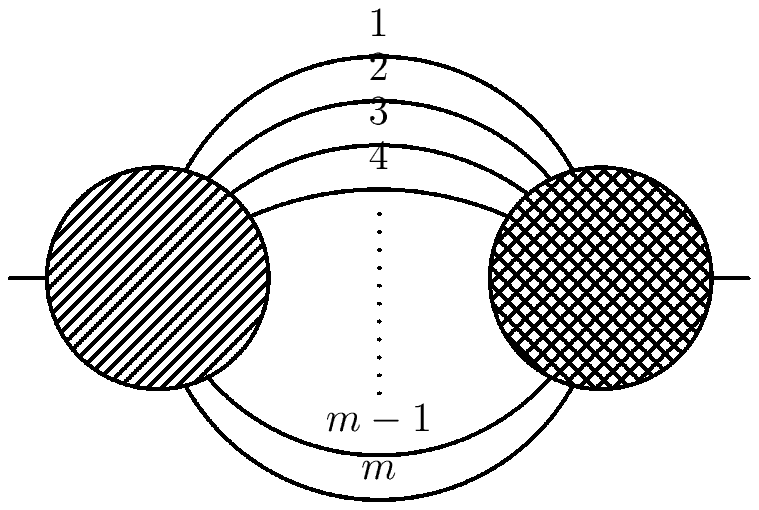,width=6.0cm}}
\caption{A self-energy diagram at any order in the coupling can always
be rearranged in this form with momentum $p_i$ flowing through line $i$
except the last line $m$ where the momentum $p_m = k+\sum_i^{m-1} p_i$. 
The two blobs may be complex vertices or $(m+1)$-point Green's functions
and they need not be the same, hence the different shadings. There 
are of course more than one possibility of such an arrangement for 
each graph.} 
\label{f:se}
\efi

\subsection{The Case of All Bosonic External Lines}
\label{s:bos}

Our own approach to arrive at the results in \eforef{eq:pb}{eq:pf}{eq:db}{eq:df} 
is by using the imaginary-time formalism and performing the contour integrals
of the loop momenta $p_1, p_2 \dots p_{m-1}$, without loss of generality,
in that order. Furthermore, lines $1$ to $m$ will all be taken to be bosonic 
and massless for simplicity. The case that some lines are bosonic and 
some fermionic is a generalization which does not affect the proof 
provided equations to be used below are suitably modified. This will be 
touched upon briefly in Sect. \ref{s:mix}. Since the discontinuity of the 
self-energy contains a number of different processes, we will aim 
at getting only one process, say that with the overall energy-momentum 
delta function 
\be \d^{(4)}(k+\sum_i^m s_i p_i)  
\label{eq:delta}
\ee
where the $s_i$'s is a fixed set of signs associated with this 
particular process depending on which momentum $p_i$ is outgoing or 
incoming. We adopt the convention that $s=+$ for incoming or absorption
of a particle and $s=-$ for outgoing or emission. This applies also 
to $k$. By concentrating on \eref{eq:delta}, any pole in the contour 
integrations that does not contribute to this particular chosen 
process need not be picked up and will be dropped from the 
discussion. For each line, which one of the two poles can 
be picked is determined by the signs $s_i$'s so that there is 
only one pole per line that will contribute to \eref{eq:delta}. 
Starting from the $p^0_1$ contour integration, it can either pick the
pole on line 1 or that on line $m$. In either case, the $p^0_2$ integration
will have also two possibilities, that of line 2 or line $m$ in the former
case and line 2 or line 1 in the latter. Since each contour integration
will have two possibilities to pick a pole so that there will be 
$2^{m-1}$ terms that contribute to \eref{eq:delta} after all $m-1$ contour 
integrations have been performed. Picking a pole on a line in a contour
integral is equivalent to putting that line on-shell and giving it a 
factor of a thermal distribution. Since there are $m-1$ integrations and 
$m$ lines, all but one line will remain off-shell at the end. This line 
will be cut and put on-shell as well when the discontinuity of the self-energy 
is finally taken. Therefore although in general the numerators of the 
$2^{m-1}$ terms will be different after the integrations because different 
sets of external lines may have been put on-shell, they will become 
identical once the discontinuity has been taken. We take the numerator 
$\cn(k,p_1,p_2,\dots,p_{m-1})$ as the part that does not include the thermal 
distributions and is evaluated at the mass-shells of these particles.
Because it is the same for every term, it can be taken out of the 
following discussion as a common factor. We can concentrate entirely on 
manipulating the products of thermal distributions into the desired 
form. Those readers not interested in the details of this manipulation
could accept on faith that the final form of the product of 
distributions is correct and skip to section B to only examine the
identities and the simple examples in section C. 

After the discontinuity has been taken, there will be $2^{m-1}$ terms,
each with a different product of thermal distributions. Writing the 
propagators in the convenient form
\be   \frac{1}{p^2_{0i}-p^2_i} 
    = \frac{1}{2 p_i} \sum_{s_i=\pm} \frac{s_i}{p_{0i}-s_i p_i}
\ee
for $i=1$ to $m-1$ and
\be   \frac{1}{p^2_{0m}-p^2_m}
    = \frac{1}{2 p_m} \sum_{s_m=\pm} \frac{-s_m}{p_{0m}+s_m p_m} \; ,
\ee
where $p_i =|\P_i|$ for $i=1,\dots,m-1$ and $p_m = |\K+\sum_i^{m-1} \P_i|$.
If the $p^0_1$ contour integration picks up the pole on line 1, there will 
be the factor $-(1/2+\fp(s_1 p_1))$ which includes also the vacuum part.  
On the other hand if it is the pole on line $m$ that is picked, then there 
will instead be $(1/2+\fp(-s_m p_m))$. In the first case, subsequent 
integration of $p_2^0$ will yield either the $-(1/2+\fp(s_2 p_2))$ 
or $(1/2+\fp(-s_1 p_1-s_m p_m))$ thermal factor depending on whether 
the pole on line 2 or line $m$ is picked. In the other case, there 
will be $-(1/2+\fp(s_2 p_2))$ or $(1/2+\fp(-s_1 p_1-s_m p_m))$ 
from line 2 and line 1 respectively. Continuing like so, there will be 
different products of thermal distributions. Starting from the case 
where each contour integration picks the pole of its own line, then only 
one of the lines $1$ to $m-1$ picks the pole of line $m$ instead of its
own line, next two lines do not pick their own lines and so on until none 
of the $p_i^0$ contour integration pick their own pole at $p_i^0 = s_i p_i$. 
Using the simplifying notations, 
\be \fp_i    = \fp(s_i p_i)   \; ,
\label{eq:pgtr}
\ee 
\be \fp_{-j} = \fp(-s_j p_j)  \; ,
\ee 
\be \fp_{ m+i+j+\cdots} = \fp(+s_m p_m+s_i p_i+s_j p_j+\cdots) \; ,
\ee
\be \fp_{-m-i-j-\cdots} = \fp(-s_m p_m-s_i p_i-s_j p_j-\cdots) \; ,
\ee
and
\be \tfp_i = 1/2 +\fp_i  
\ee
etc. 
The sum of thermal factors from the $2^{m-1}$ terms with all common
factors taken out can therefore be written as 
\bea \eF &&=\; \prod_i^{m-1} \tfp_i                             
        -   \sum_{j=1}^{m-1} 
             \Big ( \prod_{i\not =j}^{m-1} \tfp_i \Big )
                    \tfp_{-m-\sum_{i\not =j}^{j-1} i}            \nonumber \\
        +&& \sum_{j=1}^{m-2} \sum_{k=j+1}^{m-1} 
             \Big ( \prod_{i\not =j\not =k}^{m-1} \tfp_i \Big )
                    \tfp_{-m-\sum_{i\not =j}^{j-1} i}           
             \tfp_{-m-\sum_{i=1}^{k-1} i}                        \nonumber \\  
        -&& \sum_{j=1}^{m-3} \sum_{k=j+1}^{m-2} \sum_{l=k+1}^{m-1} 
             \Big ( \prod_{i\not =j\not =k\not = l}^{m-1} \tfp_i \Big )
                    \tfp_{-m-\sum_{i\not =j}^{j-1} i}            \nonumber \\  
         && \mbox{\hspace{2.0cm}}
      \times \tfp_{-m-\sum_{i=1}^{k-1} i} \tfp_{-m-\sum_{i=1}^{l-1} i} 
                                                                \nonumber \\  
        +&& \;\; \cdots \cdots \cdots \cdots \cdots \cdots      \nonumber \\
        +&& (-1)^{m-1} \tfp_{-m} \tfp_{-m-1} \tfp_{-m-1-2} \tfp_{-m-1-2-3} 
            \cdots \cdots \cdots                                \nonumber \\
         && \;\; \times \tfp_{-m-1-2-\cdots-(m-2)}  \; .
\label{eq:thf}
\eea
The key to finding a way out of this seemingly endless sum of terms
is the observation that each term must have a partner. By that we
mean for every term containing $\tfp_{m-1}$ as part of the thermal weight, 
there must be another one that differs from this term only in this factor 
by having $-\tfp_{-m-1-2-\cdots-(m-2)}$ in its place instead. This is 
true because no matter which sets of poles were picked in the contour 
integrations, the last integration of $p^0_{m-1}$ must be able to pick 
either its own pole on line $m-1$ or the pole on the other line that now 
carries the energy $p^0_{m-1}$. This could be the line $m$ if it has not
yet been touched or another line with the energy $p^0_{m-1}$
shifted there via an already performed contour integration. Therefore
every term must have either $\tfp_{m-1}$ or $-\tfp_{-m-1-2-\cdots-(m-2)}$
by virtue of the last contour integration $p^0_{m-1}$. They can thus all be 
paired. Now using the delta function \eref{eq:delta}, we get the identity
\bea & &   \tfp_{m-1}- \tfp_{-m-1-2-\cdots-(m-2)}               \nonumber \\ 
     & & = \fp_{m-1} - \fp_{-m-1-2-\cdots-(m-2)}                \nonumber \\ 
     & & = \big ( e^{-k^0/T}-1 \big ) \; \fp_{m-1} \fp_{m+1+2+\cdots+(m-2)} \; .
\label{eq:id1}
\eea
After this is applied to every pair, the common factor on the r.h.s. of
\eref{eq:id1} can be taken out of \eref{eq:thf}. We now have $2^{m-2}$
terms to sort out. 

Shifting focus now onto the thermal weight $\tfp_{m-2}$ or 
$-\tfp_{-m-1-2-\cdots-(m-3)}$. For similar reason as before, each term 
must have either one of the two and these again form pairs because of the 
second last contour integration of $p^0_{m-2}$. The identity 
\bea & & \Big (\fp_{m-2} - \fp_{-m-1-2-\cdots-(m-3)} \Big ) 
         \fp_{m+1+2+\cdots+(m-2)}                               \nonumber \\ 
     & & = \fp_{m-2} \fp_{m+1+2+\cdots+(m-3)} 
\label{eq:id2}
\eea 
is a more general form of \eref{eq:id1} and can be used to remove the
last factor in the now common thermal factor on the r.h.s. of 
\eref{eq:id1} to get the new common factor
\be \big ( e^{-k^0/T}-1 \big ) \; \fp_{m-1} \; \fp_{m-2} 
    \; \fp_{m+1+2+\cdots+(m-3)}   \; .                              
\ee
The number of terms has now been further reduced to $2^{m-3}$. 
Iterating this thermal factor reduction process, all $2^{m-1}$ terms 
can be grouped together eventually into one common thermal factor 
\bea \eF &=& \big ( e^{-k^0/T}-1 \big ) \; \fp_m \; \fp_{m-1} \; \fp_{m-2} 
             \; \fp_{m-3} \; \cdots \cdots                      \nonumber \\
         & & \times \cdots \cdots \fp_3 \fp_2 \fp_1  \; . 
\eea
The first factor will be divided out by the denominator in
\eref{eq:pb} with $k^0 \lra -k^0$ for production or that in \eref{eq:db}
for decay. The remaining thermal distributions will be $\fp(p)$ 
for absorption with $s=+$ in the delta function in \eref{eq:delta}
or $-(1+\fp(p))$ for emission when $s=-$ as expected. Combining 
the sign factors $s_1 s_2 \cdots s_m$ originating from the propagators 
in \eref{eq:pgtr} but have been left out of the discussion so far, 
this is the $F_\cp$ in \etwref{eq:i}{eq:distr}
\be  F_\cp = \frac{s_1 s_2 \cdots s_m}{e^{-k^0/T}-1} \; \eF  \; .
\ee

Now we turn to the phase space integrals in \eref{eq:i}. From the
$m-1$ loop integrations, there are already the 3-momentum measures 
${d^3 p_i}/{(2\p)^3}$ for $i=1$ to $m-1$. This together with the 
$1/(2p_i)$ from the propagators give $d\Phi$ in \eref{eq:psm} from
$i=1$ up to the $m-1$ entry. The last entry $m$ can be gotten by 
introducing the identity 
\be \int \frac{d^3 p_m}{(2\p)^3} \; (2\p)^3 \; 
    \d^{(3)} \big (k+\sum^m_{i=1} s_i p_i \big ) = 1    \; . 
\ee
Multiplying this by the remaining $1/2p_m$ and the energy delta function 
$(2\p) \d (k^0+\sum^m_{i=1} p^0_i)$ from the discontinuity gives  
\be \int \frac{d^4 p_m}{(2\p)^4} \; (2\p) \d^{(s_m)} (p^2_m) 
    \; \d^{(4)} \big (k+\sum^m_{i=1} s_i p_i \big )     \; . 
\ee 
By grouping everything else not discussed above which is mainly 
the numerator $\cn(k,p_1,p_2,\cdots,p_m)$ together, this will be 
$|\cm_\cp|^2$ if in this particular process the two blobs in \fref{f:se} 
contain no internal loop momentum that runs completely inside the blob. 
If either one or both contain internal loops then this will be the product 
of $(\bar \cm^{}_{\cp} \cm^*_{\cp} +\bar \cm^*_{\cp} \cm^{}_{\cp})$ 
and $\cf_{\cp}$ in the function $J(k)$. The latter $\cf_{\cp}$ factor
will carry essentially the hidden loop momentum integrations and there 
will be associated thermal weights. We will however adopt an approach in 
which there will not be any loop. But this and the remaining 
parts of the function $J(k)$ are the subjects of Sect. \ref{s:is}.

\subsection{The Case of a Mixture of Bosonic and Fermionic External Lines}
\label{s:mix}

For the more general case where there is a mixture of fermionic and 
bosonic lines amongst the lines $1$ to $m$ in \fref{f:se}, the proof
is somewhat more complicated because bosonic distribution can turn
into fermionic distribution and vice versa when they are evaluated
at a pole depending on whether this has an imaginary part with a 
total of an integer $n$ or a half-integer $n+1/2$ of $(2\p i)$ imaginary
energy. For a bosonic self-energy, there must be an even number of
intermediate fermion lines and for a fermion self-energy, there must be 
an odd number of such lines. In the latter case in order not to have a 
contradiction in our momentum arrangement in \fref{f:se}, the last
line $m$ must be fermionic whereas in the bosonic self-energy
there is no such constraint. We could nevertheless arrange all
fermion lines to be those at the bottom and leaving all boson lines 
at the top in either case. That is if there are $n$ boson lines 
$n\le m$, we arrange the lines so that line 1 to $n$ will be bosonic 
and line $n+1$ to $m$ will be fermionic. Then for a fermion 
self-energy the last contour integration will give the pair combination
\bea & &   -\tfm_{m-1}+\tfp_{-m-1-2-\cdots-(m-2)}             \nonumber \\
     & & =  \fm_{m-1} +\fp_{-m-1-2-\cdots-(m-2)}              \nonumber \\
     & & = -\big ( e^{-k^0/T}+1 \big ) \;  \fm_{m-1} \; \fp_{m+1+2+\cdots+(m-2)}  
\eea
if the line $m-1$ is fermionic or 
\bea & &   -\tfp_{m-1}+\tfm_{-m-1-2-\cdots-(m-2)}             \nonumber \\
     & & = -\fp_{m-1} -\fm_{-m-1-2-\cdots-(m-2)}              \nonumber \\
     & & = -\big ( e^{-k^0/T}+1 \big ) \;  \fp_{m-1} \; \fm_{m+1+2+\cdots+(m-2)}  
\eea
if this line is bosonic. This is always true because the number of
fermion lines connecting the two blobs in \fref{f:se} is odd. 
Here we have used the notation $\tfm_i = 1/2-\fm_i$. 
For a bosonic self-energy, there is the combination 
\bea & &    \tfm_{m-1}-\tfm_{-m-1-2-\cdots-(m-2)}             \nonumber \\
     & & = -\fm_{m-1} +\fm_{-m-1-2-\cdots-(m-2)}              \nonumber \\
     & & =  \big ( e^{-k^0/T}-1 \big ) \;  \fm_{m-1} \; \fm_{m+1+2+\cdots+(m-2)}  
     \; .
\eea
The thermal factors can be sorted into the desire form in
\eref{eq:distr} using the identities
\bea -\Big (\fm_i - \fm_{-j} \Big ) \;\fp_{i+j} &=& \fm_i \fm_j \;,     \\
      \Big (\fp_i + \fm_{-j} \Big ) \;\fm_{i+j} &=& \fp_i \fm_j \;,     \\
     -\Big (\fm_i + \fp_{-j} \Big ) \;\fm_{i+j} &=& \fm_i \fp_j            
\eea
iteratively as similarly done before in the previous section. The phase 
space integration measure and the rest are the same as discussed 
there. The proof of the main form of the function $I(k)$ or $J(k)$ is 
therefore complete. In Sect. \ref{s:is}, the integrand and substructures 
in $I(k)$ and $J(k)$ or equivalently the internal structures of the blobs 
in \fref{f:se} will be the subject.

\subsection{Examples}
\label{s:ex_distr}

In this subsection, we give two examples of the thermal factor 
reduction from $2^{m-1}$ terms down to one discussed in the previous
section. First we consider a bosonic self-energy with $m=3$ and 
one-fermion loop. In accordance to our discussions above, we push
all fermion lines to the bottom so line 1 will be boson and 
line 2 and 3 will be fermion. Without stating explicitly what
process to get from the self-energy, we choose the very general 
energy conserving delta function $\d(k_0+s_1 p_1 +s_2 p_2 +s_3 p_3)$. 
After the discontinuity has been taken, the sum of thermal distributions
is
\bea F'= \;\;\; s_1 s_2 s_3 && 
         \Big (-\tfp_1 \tfm_2      +\tfm_{-3} \tfm_2                   \nonumber \\
      && \;\;\;\;  +\tfp_1 \tfm_{-1-3} -\tfm_{-3} \tfm_{-1-3} \Big )   \; .
\eea  
This can be reduced by pairing as follows. 
\bea F'= \;\;\; s_1 s_2 s_3 && 
          \Big (\tfp_1-\tfm_{-3} \Big )\Big (-\tfm_2+\tfm_{-1-3} \Big ) \nonumber \\
       = \;\;\; s_1 s_2 s_3 && 
          \Big (\fp_1+\fm_{-3} \Big )\Big (\fm_2-\fm_{-1-3} \Big )      \nonumber \\
       = -  s_1 s_2 s_3 && \fp_1 \fm_2 \fm_3 \big (e^{-k_0/T}-1 \big ) 
\eea

As another example, we consider a fermion self-energy with $m=4$. 
This time only line $m$ is fermionic. The sum of thermal distributions 
now is 
\bea F'= s_1 s_2 s_3 s_4 && 
         \Big ( -\tfp_1 \tfp_2 \tfp_3 +\tfm_{-4} \tfp_2 \tfp_3            \nonumber \\
      &&  +\tfp_1 \tfm_{-1-4} \tfp_3 +\tfp_1 \tfp_2 \tfm_{-1-2-4} \nonumber \\
      &&  -\tfm_{-4} \tfm_{-1-4} \tfp_3 -\tfm_{-4} \tfp_2 \tfm_{-1-2-4}
                                                                          \nonumber \\
      &&  -\tfp_1 \tfm_{-1-4} \tfm_{-1-2-4} 
          +\tfm_{-4} \tfm_{-1-4} \tfm_{-1-2-4}  \Big ) .                  \nonumber \\
\eea
Pairing as before gives
\bea F'= s_1 s_2 s_3 s_4 && 
         \Big (\tfp_1-\tfm_{-4} \Big )\Big (\tfp_2-\tfm_{-1-4} \Big )     \nonumber \\ 
      && \times \Big (\tfp_3-\tfm_{-1-2-4} \Big )                         \nonumber \\ 
       = s_1 s_2 s_3 s_4 && 
         \Big (\fp_1+\fm_{-4} \Big )\Big (\fp_2+\fm_{-1-4} \Big )         \nonumber \\ 
      && \times \Big (\fp_3+\fm_{-1-2-4} \Big )                           \nonumber \\ 
       = s_1 s_2 s_3 s_4 && \fp_1 \fp_2 \fp_3 \fm_4 \big (e^{-k_0/T}-1 \big ) \; .
\eea

\section{The Internal Structures of \\ I(k) and J(k)} 
\label{s:is}

The previous sections showed how the discontinuity of the self-energy
could be arranged into phase space integrations over all the external 
lines weighed by their respective thermal distributions. There we called 
any line that showed itself explicitly in \fref{f:se} external and these 
all have entries in the overall energy-momentum conserving delta function
in \eref{eq:i} and \eref{eq:j}. In this section, we will deal with the 
remaining structures in these equations. These structures come essentially 
from the blobs in \fref{f:se} which have not been discussed yet. 

In the simplest case, each blob consists of a tree graph with all 
lines leading to the external lines 1 to $m$ and the line with 4-momentum 
$k$. This can be just a few lines at low orders or a very large tree at high 
orders. All internal lines' energies-momenta are fixed completely by the 
external lines and there is no additional integration and thermal distribution
other than those of the external lines. If the two blobs are identical,
then there is automatically the $|\cm_\cp|^2$. If they are not, then there
is the interference $\bar \cm^{}_{\cp} \cm^*_{\cp} +\bar \cm^*_{\cp} \cm^{}_{\cp}$
with $\cf_\cp =1$. In the case that $\bar \cm^{}_{\cp}$ and $\cm^{}_{\cp}$
are of the same order in the coupling, there must also be the possibilities  
$|\cm^{}_{\cp}|^2$ and $|\bar \cm^{}_{\cp}|^2$ so that a larger squared
modulus can be formed $|\bar \cm^{}_{\cp}+\cm^{}_{\cp}|^2$. For familiar
field theories such as $\f^3$, $\f^4$, QED, QCD etc., tree graphs with
the same number of external legs are of the same order so this grouping
into a larger squared modulus is always possible. This case for $J(k)$
is rather trivial but has to be stated first for the sake of completeness 
before we move on to more complex cases. So non-identical blobs with no 
internal loop are the simplest examples of interference graphs in $J(k)$ 
that can actually be grouped into the class of function $I(k)$. It is the 
more complex interference graphs genuinely belonged to the $J(k)$ 
function part that are our main concern. Towards this we now turn. 

In the previous paragraph, we discussed the case of the tree graphs. 
This type of contributions can generally be interpreted straight 
forwardly as scattering processes or other interactions. For cases  
of a few particles in the initial state, there will always be the 
vacuum counterpart. These are quite familiar physically. We need not 
say anymore on these. What we find less obvious and less well-known is
the physical interpretation of the case that the blobs \fref{f:se} 
have internal loops. Within the imaginary-time formalism, 
it is not too difficult, albeit tedious, to perform the calculation of 
these graphs and it is all too easy to lose the feel of what are the
actual physical processes involved in any particular calculation. 
The nature of these graphs made it even harder because of the fact 
that first there is no vacuum counterpart and second they are the 
non-intuitive interference type. If one is not clear in one's thinking
and clings on to the over-familiar vacuum picture, they could even 
appear to be impossible. 

\bfi
\centerline{
\epsfig{figure=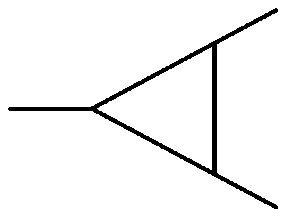,width=2.0cm}}
\caption{An example graph for a blob with one loop.} 
\label{f:emabega}
\efi

Let us consider the simple case that one blob has one internal loop and 
the other none, the procedure in arriving at the form of $I(k)$ has 
already put all the external lines on-shell. So the calculation of this 
blob is equivalent to the evaluation of a one-loop $n$-point function 
where $n \le (m+1)$. The rules of the imaginary-time formalism stipulated 
that the energy contour integration of the loop must pick out every pole 
in the propagators in turn of all the particles along the lines that form 
the loop. This step is equivalent to putting in turn every particle 
propagating around the loop on-shell. For each particle around the
loop, there are two poles. One can be interpreted as its emission into
the heat bath from one vertex and its absorption from the bath by another. 
The other is absorption from the heat bath now at the first vertex
and its emission back into it from the second. An example graph
for a blob with one internal loop and three external lines is shown in 
\fref{f:emabega}. This graph can be turned into the six graphs in 
\fref{f:emab} after the energy contour integration put the three internal
lines in turn on mass-shell. These are indicated by the now on-shell large 
dotted lines in \fref{f:emab}. 

\bfi
\centerline{
\epsfig{figure=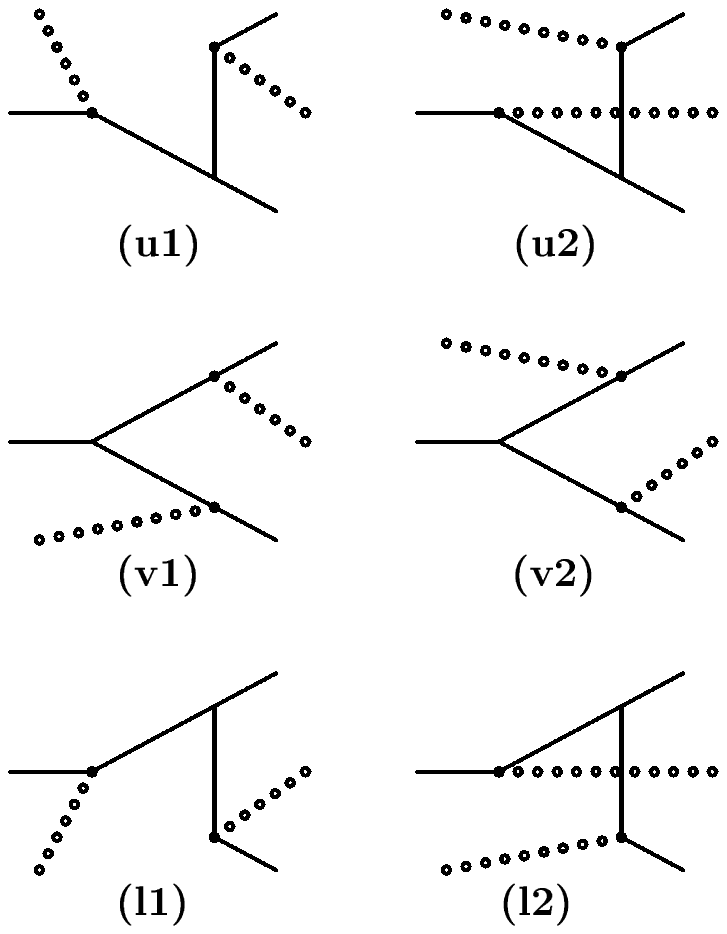,width=4.50cm}}
\null
\caption{Performing the energy contour integration is equivalent to putting
the internal lines in turn on-shell. (u1) and (u2), (v1) and (v2), and 
(l1) and (l2) are graphs resulting from putting the upper, vertical and 
lower internal line respectively on-shell. Each pair of graphs corresponds 
to the two poles from each line.} 
\label{f:emab}
\efi

In our special case, it is easy to turn the blob with the loop into
a sum of tree graphs with emission-absorption of particles of the same 
momentum. If one now takes a step back and looks again at the whole 
picture, there is also the other blob that helps make up the whole 
self-energy which consists of only a tree graph. That for our example 
graph in \fref{f:emabega} will be that shown in \fref{f:emabegb}. 

\bfi
\centerline{
\epsfig{figure=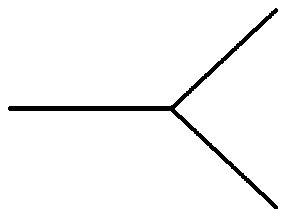,width=2.0cm}}
\caption{The associated graph to \fref{f:emabega}.}
\label{f:emabegb}
\efi

Remembering that \ftwref{f:emabega}{f:emabegb} together form an
interference contribution, opening up the loop in \fref{f:emabega} 
would seem to render the interference impossible because there
must be the same incoming as well as outgoing particles in both 
graphs. However, graphs representing interactions in a heat 
bath are not quite the same as those in the vacuum. It is
a common practice to use the same graphs in both situations but it
must not be forgotten that in a heat bath graphs should be understood
to be implicitly accompanied by spectator particles that made up the
multiparticle system. So opening up the loop in \fref{f:emabega} 
should be accompanied by a change of \fref{f:emabegb} into 
\fref{f:emabegc}. The momentum carried by the spectator line in 
\fref{f:emabegc} is of course different when this is convoluted
with each of the three (u), (v) and (l) pairs in \fref{f:emab}. 

\bfi
\centerline{
\epsfig{figure=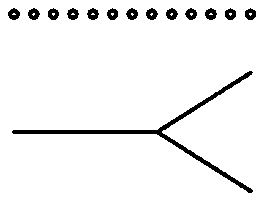,width=2.0cm}}
\caption{This is how \fref{f:emabegb} should be properly represented
once the loop in \fref{f:emabega} has been opened up.} 
\label{f:emabegc}
\efi

So in this simple example, there will be six terms in the function
$J(k)$. Each will have a phase space integration of either $s=\pm$ form 
\be   d \mbox{\boldmath $\varphi$}_{\cp'}
    = \frac{d^4 l}{(2 \p)^4} \; (2 \p) \d^{(s)} (l^2-m^2) 
\ee
originating from the previous loop integration. Because these
correspond to the emission and absorption of particles of the same
momentum within a single graph, they do not play a part in the 
overall energy-momentum flow and have not an entry in the 
corresponding constraining delta function. The emission and
absorption of these particles must come with thermal distribution. 
Because of the unusual nature of the emission-absorption or the 
absorption-emission in the same graph, each of these in our example
with only one internal loop acquires the distribution 
\be \cf_{\cp'} = 1/2 + \fp(|l^0|) 
\ee
for boson or
\be \cf_{\cp'} = 1/2 - \fm(|l^0|) 
\ee
for fermion in this simple case. For more complex cases where the
blobs are not identical and each has several internal loops, after opening
up each loop in the blobs and adding the corresponding spectator on the 
other blob the phase space integration becomes 
\be   d \mbox{\boldmath $\varphi$}_{\cp'}
    = \prod^L_i \frac{d^4 l_i}{(2 \p)^4} \; (2 \p) \d^{(s_i)} (l^2_i-m^2_i) 
      \; ,
\ee
where $L$ is the combined total number of internal loops between the
two blobs. If there are any self-energy insertions hidden in a blob,
then this loop will have to be opened up too. Therefore each self-energy
insertion will be turned into the emission and absorption of a particle 
of the same 4-momentum from a single line. If the line is fermionic,
this is the usual forward scattering, but there is also the case that
a bosonic line emits and absorbs or vice versa a fermion of the same
4-momentum. Thus we label these generically as generalized forward 
scattering. In this case, the internal loop turned phase space 
integrations will be of the form of \eref{eq:phint}. 
The factor of thermal distribution when there are many loops becomes
\be \cf_{\cp'} = \prod^L_i \big (1/2+ \es_i f^{(\ess_i)} (|l^0|) \big ) \; .
\label{eq:ldistr}
\ee
Because there must be four or a larger even number of graphs obtainable 
from each loop, the sum in \eref{eq:j} over $\cp'$ is a sum over the 
main processes as well as over the possible subprocesses of 
emission-absorption of particles in each main process. 

The steps in arriving at the form of the $J(k)$ part contribution 
to the production from or the decay of a particle in a heat bath 
are now shown. So expressing a production or a stimulated 
decay of a particle in a heat bath in terms of all the contributing 
processes should not be restricted only to processes expressible in
terms of squared modulus amplitudes. In other words, there is the very
important $J(k)$ interference contributions in addition to the better
known $I(k)$ part. In the next section we will illustrate all these in 
an explicit example. But before we do that it must be mentioned
the vacuum parts still have to be regularized and renormalized 
in the usual way. We have assumed that this was implicitly understood.

\section{A More Complex And Complete Example}
\label{s:ceg}

In this section we will use QED coupled to a massive vector particle 
as an example. The relevant lagrangian is
\bea \cl &=& -\frac{1}{4} F^{\m\n} F_{\m\n} 
             +\bar \j \, \g^\m (i\del_\m -e A_\m -g V_\m)\, \j    \\ \nonumber
         & & -\frac{1}{4} G^{\m\n} G_{\m\n} 
             -\frac{1}{2} M^2\, V^\m V_\m \; , 
\label{eq:lag}
\eea 
where $F^{\m\n}$ and $G^{\m\n}$ are the field tensors for the $A^\m$
and $V^\m$ vector fields respectively. The coupling $g$ is taken to be
much weaker than $e$, $e\gg g$, so only higher order corrections in $e$ will 
be considered. Our heat bath will consist only of leptons and photons at a 
temperature $T$, and the vector $V^\m$ is not itself thermalized. A massive 
vector with a mass $M \gg T$ will be sent into the heat bath to determine 
the medium modification of its width up to two loops. 

\bfi
\centerline{
\epsfig{figure=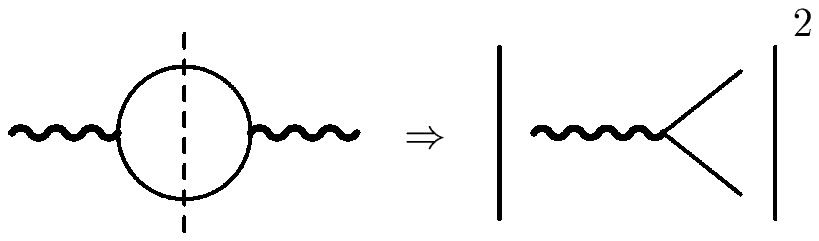,width=6.5cm}}
\caption{The leading contribution to the decay of $V^\m$.}
\label{f:1lse}
\efi

At leading order (LO), there is only the decay into a dilepton pair 
contribution to \eref{eq:db} coming from the one-loop self-energy
shown in \fref{f:1lse}. This is of course only a contribution of the
type of $I(k)$. From \eref{eq:gam_ol} after some substitutions, this 
contribution is  
\bea && 2 M \; \Gamma \big |_{\mbox{\scriptsize LO}}                \nonumber \\
       =\int && \frac{d^4 p}{(2\p)^4}  (2\p) \d^{(+)} (p^2)      
                \frac{d^4 p'}{(2\p)^4} (2\p) \d^{(+)} (p'^2)        \nonumber \\ 
      \times && (2\p)^4 \d^{(4)} (k-p-p')                           \nonumber \\ 
      \times && \big (1-\fm(p_0) \big )\big (1-\fm(p_0') \big) 
                \; |\cm|^2_{\mbox{\scriptsize LO}}   \; . 
\label{eq:lo}
\eea
Here the numerator is just the familiar Dirac trace of the lepton loop,
one gets after summing over final spins and averaged over initial spin
\be   |\cm|^2_{\mbox{\scriptsize LO}} 
    = \frac{4 g^2}{3} \Big \{ p\cdot p'+\frac{2(p\cdot k)(p'\cdot k)}{M^2} \Big \}\;.
\ee

At the next-to-leading order, there are only three self-energy graphs
but there are many physical processes hidden within these graphs.
In the following graphs, the thick wavy line is for the massive vector 
while the thin wavy line is for photon. The lines overlaid by the 
vertical dashed line are the exposed external lines of \fref{f:se}. 
First we examine those arrangements of these graphs with no 
internal loops within the blobs. By putting the three internal lines 
in the two-loop graph with an internal self-energy in \fref{f:2lses1} 
on-shell, three physical processes emerge. We get a sum of the 
squared modulus amplitudes of two compton scattering in \fref{f:comps}, 
of one decay with photon emission in \fref{f:decs} and of one vector-photon 
fusion in \fref{f:fuss}. These are clearly all contributions to the 
$I(k)$ function part in \eref{eq:db}. There will of course be the
other contributions where the internal self-energy is on the lower 
lepton line. These can easily be taken care of by a simple factor 
of two by symmetry. 
\bfi
\centerline{
\epsfig{figure=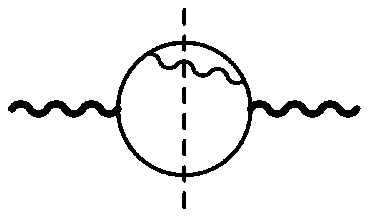,width=3.0cm}}
\caption{Two-loop self-energy with internal self-energy insertion. 
All three internal lines in the middle will be put on-shell as external
lines therefore the two blobs are tree graphs.}
\label{f:2lses1}
\null
\centerline{
\epsfig{figure=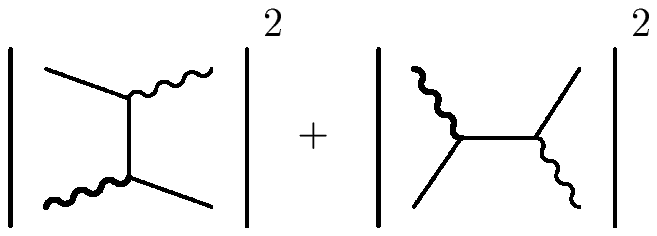,width=5.0cm}}
\caption{Contributions to compton scattering from \fref{f:2lses1}.}
\label{f:comps}
\null
\centerline{
\epsfig{figure=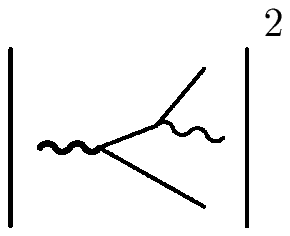,width=2.5cm}}
\caption{Decay with photon emission from \fref{f:2lses1}.}
\label{f:decs}
\null
\centerline{
\epsfig{figure=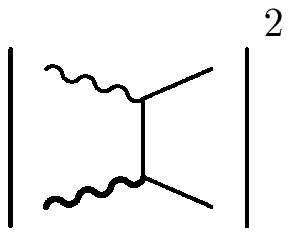,width=2.5cm}}
\caption{Vector-photon fusion contribution from \fref{f:2lses1}.}
\label{f:fuss}
\efi

The other two-loop graph is the one with a vertex correction drawn 
in \fref{f:2lsev1}. Putting all three internal lines on-shell, three
physical, albeit interference, processes emerge. They are again
compton scattering in \fref{f:compv}, three-body decay in \fref{f:decv}
and fusion in \fref{f:fusv}. Because they are all interference
contributions, it appears that they belong to $J(k)$. As we said in
Sect. \ref{s:is}, blobs that have no internal loop or have only a 
tree structure could be regrouped with other contributions to form a 
larger amplitude. Examining the pairs in \ftwref{f:comps}{f:compv}, 
\ftwref{f:decs}{f:decv} and \ftwref{f:fuss}{f:fusv} and also not 
forgetting that the two other lepton lines together with 
the photon line in \fref{f:2lsev1} could also be put on-shell as well 
to give similar contributions, this can of course be done as 
is well-known. We merely broke down the contributions into parts so 
that they could be clearly seen, within the framework that we are 
presenting the paper, which contribution came from which diagram. Loosely 
speaking the contributions from \fref{f:2lsev1} should be in $J(k)$ but 
it is better to tighten the definition so that this function is 
restricted to the genuine, not well-known in-medium interference 
contributions to the modification of the width.   
\bfi
\centerline{
\epsfig{figure=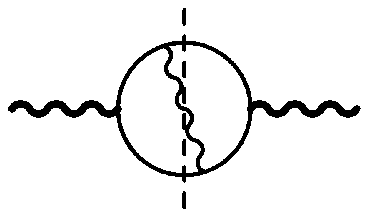,width=3.0cm}}
\caption{Two-loop self-energy with vertex correction. Putting the
three intermediate lines on-shell and there will not be any internal
loop.} 
\label{f:2lsev1}
\null
\centerline{
\epsfig{figure=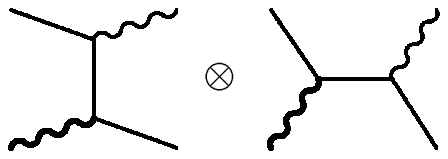,width=4.0cm}}
\caption{Interference contribution to compton scattering from \fref{f:2lsev1}.
This can be regrouped with \fref{f:comps} to form a larger squared modulus
of a single amplitude. It does not genuinely belong to the $J(k)$ class.} 
\label{f:compv}
\null
\centerline{
\epsfig{figure=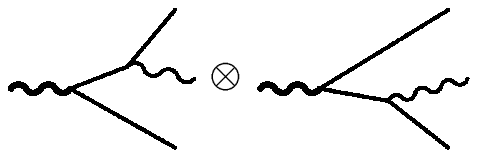,width=4.0cm}}
\caption{Interference contribution to decay with photon emission 
from \fref{f:2lsev1}. This can also be regrouped with \fref{f:decs} to 
form a larger squared modulus of a single amplitude. This is not 
really a $J(k)$ class contribution.}
\label{f:decv}
\null
\centerline{
\epsfig{figure=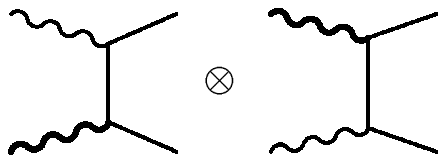,width=4.0cm}}
\caption{Interference vector-photon fusion contribution from \fref{f:2lsev1}.
Again together with \fref{f:fuss} this can form a large squared modulus of 
a single amplitude. Therefore it is in the $I(k)$ class contribution.}
\label{f:fusv}
\efi

After combining the contributions from \ftwref{f:2lses1}{f:2lsev1}, 
we get for the three processes 
\bea && 2 M\; \Gamma \big |_{\mbox{\scriptsize compton}}            \nonumber \\
      =2\int && \frac{d^4 p}{(2\p)^4}  (2\p) \d^{(+)} (p^2)      
                \frac{d^4 p'}{(2\p)^4} (2\p) \d^{(+)} (p'^2)        \nonumber \\ 
      \times && \frac{d^4 q}{(2\p)^4}  (2\p) \d^{(+)} (q^2) 
                \; (2\p)^4 \d^{(4)} (k+p-p'-q)                      \nonumber \\ 
      \times && \fm(p_0) \big(1-\fm(p_0') \big)\big(1+\fp(q_0) \big) 
                                                                    \nonumber \\
      \times && |\cm |^2_{\mbox{\scriptsize compton}}  \; ,              
\label{eq:com}
\eea
\bea && 2 M\; \Gamma \big |_{\mbox{\scriptsize decay}}              \nonumber \\
       =\int && \frac{d^4 p}{(2\p)^4}  (2\p) \d^{(+)} (p^2)      
                \frac{d^4 p'}{(2\p)^4} (2\p) \d^{(+)} (p'^2)        \nonumber \\ 
      \times && \frac{d^4 q}{(2\p)^4}  (2\p) \d^{(+)} (q^2) 
                \; (2\p)^4 \d^{(4)} (k-p-p'-q)                      \nonumber \\ 
      \times && \big(1+\fp(q_0) \big)\big (1-\fm(p_0) \big)
                \big(1-\fm(p_0') \big)                              \nonumber \\
      \times && |\cm |^2_{\mbox{\scriptsize decay}}   \; ,
\label{eq:dec}
\eea
\bea && 2 M\; \Gamma \big |_{\mbox{\scriptsize fusion}}             \nonumber \\
      =2\int && \frac{d^4 p}{(2\p)^4}  (2\p) \d^{(+)} (p^2)      
                \frac{d^4 p'}{(2\p)^4} (2\p) \d^{(+)} (p'^2)        \nonumber \\ 
      \times && \frac{d^4 q}{(2\p)^4}  (2\p) \d^{(+)} (q^2) 
                \; (2\p)^4 \d^{(4)} (k+q-p-p')                      \nonumber \\ 
      \times && \fp(q_0) \big(1-\fm(p_0) \big)\big(1-\fm(p_0') \big) 
                                                                    \nonumber \\
      \times && |\cm |^2_{\mbox{\scriptsize fusion}}  \; .
\label{eq:fus}
\eea
The averaged over initial spins, summed over final spins matrix 
elements for the three processes are 
\bea |\cm|^2_{\mbox{\scriptsize compton}}
               &=& -\frac{4 e^2 g^2}{3}\bigg \{\frac{s}{u} + \frac{u}{s}
            +2t \left(\frac{1}{s} + \frac{1}{u} + \frac{t}{su} \right)\bigg \} 
                                                                     \nonumber \\ \\
     |\cm|^2_{\mbox{\scriptsize decay}}
               &=&  \frac{8 e^2 g^2}{3}\bigg \{ \frac{p\cdot p'+k\cdot p'} 
         {p\cdot q} + \frac{p\cdot p'+k\cdot p}{ p'\cdot q}          \nonumber \\
               & & \mbox{\hspace{1.5cm}} 
          + \frac{2(p\cdot p')^2}{(p\cdot q)( p'\cdot q)} \bigg \}             \\
     |\cm|^2_{\mbox{\scriptsize fusion}}
               &=&  \frac{4 e^2 g^2}{3}\bigg \{\frac{t}{u} + \frac{u}{t}         
            +2s \left(\frac{1}{t} + \frac{1}{u} + \frac{s}{tu} \right)\bigg \} 
             \, .                                                    \nonumber \\ 
\eea
These amplitudes were calculated in \cite{kw1} for studying the change 
in Z boson properties in the quark-gluon plasma. Note that if one is
only interested in the medium modification to the width, it is necessary
to subtract the leading order and next-to-leading order vacuum contribution 
in \etwref{eq:lo}{eq:dec}. 

Now we turn to the examples of one of the main subjects of this paper. 
In \fref{f:2lses1} instead of putting the three intermediate lines 
on-shell, one could do this to two lepton lines without touching the
photon as shown in \fref{f:2lses2}. This results in the blob on the r.h.s.
in \fref{f:se} to have an internal loop made up of two lines. As we 
discussed already in Sect. \ref{s:is}, they will be put on the mass-shell 
in turn by the energy contour integration. Thus it is possible to unfold
the diagram into several ones with emission and absorption of particles 
with the same 4-momentum within one graph. One must of course put in the 
associated spectator particles in the graph originated from the blob on 
the left-hand-side (LHS) of \fref{f:se} in order for the new graphs to make 
physical sense. These are now depicted in \ftwref{f:abems}{f:scpes}. 
In \fref{f:abems}, the massive vector decays into a real and virtual
dilepton pair. The virtual one then either absorbed a photon from and 
emitted it back into the heat bath or emitted a photon into the bath
before absorbing one back. In \fref{f:scpes}, the virtual lepton 
either annihilates with one from the heat bath to recreate another
dilepton pair or it is put on-shell via a photon exchange with a lepton
in the bath. The lepton line that has just been put on-shell is shown
in dashed line in the figure. While all these are happening on one graph, 
there is merely the vector decay in the accompanying graph. Here there 
is a photon or a lepton spectator. 

\bfi
\centerline{
\epsfig{figure=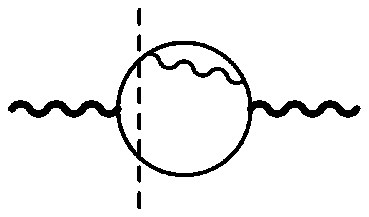,width=3.0cm}}
\caption{In this two-loop self-energy with internal self-energy insertion,
only two lines will be put on-shell. One of the blobs will have
an internal loop with two lines.}
\label{f:2lses2}
\null
\centerline{
\epsfig{figure=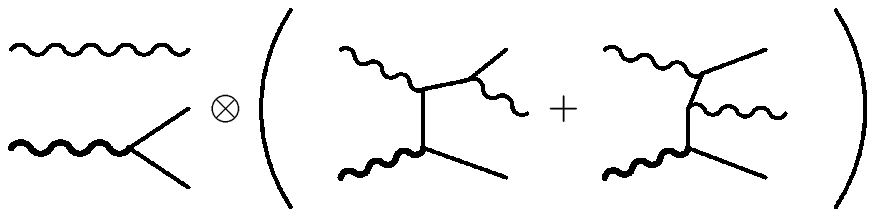,width=7.50cm}}
\caption{Two genuine $J(k)$ class contributions coming from \fref{f:2lses2}.
Photon with the same 4-momentum is emitted and absorbed within a single
graph. The same photon is required as a spectator in the other graph.}
\label{f:abems}
\null
\centerline{
\epsfig{figure=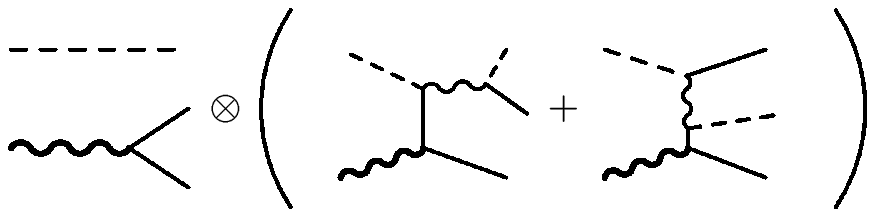,width=7.50cm}}
\caption{Opening up the fermion line in \fref{f:2lses2} results in
the absorption from and emission back of a lepton into the heat bath 
or vice versa. These are different examples of generalized forward scattering 
on the fermion line. The lepton line from the self-energy loop is now shown 
in dashed lines to distinguish them from the other leptons.} 
\label{f:scpes}
\efi

These interference graphs can be worked out to be 
\bea && 2 M \; \Gamma \big |_{\mbox{\scriptsize (\fref{f:abems})}}  \nonumber \\
      =2\int && \frac{d^4 p}{(2\p)^4} (-1)(2\p) {\d'}^{(+)} (p^2)      
                \frac{d^4 p'}{(2\p)^4} (2\p) \d^{(+)} (p'^2)        \nonumber \\ 
      \times && \frac{d^4 q}{(2\p)^4} (2\p) \d^{(+)} (q^2)
                \; (2\p)^4 \d^{(4)} (k-p-p')                        \nonumber \\ 
      \times && \big (1-\fm(p_0) \big )\big (1-\fm(p_0') \big)      \nonumber \\ 
      \times && \big (1/2+\fp(q_0) \big )
                \; (\bar \cm\;\cm^*+\bar \cm^* \cm)_{\mbox{\scriptsize (\fref{f:abems})}}   \; , 
\label{eq:abems}
\eea
and
\bea && 2 M \; \Gamma \big |_{\mbox{\scriptsize (\fref{f:scpes})}}  \nonumber \\
      =2\int && \frac{d^4 p}{(2\p)^4} (-1)(2\p) {\d'}^{(+)} (p^2)      
                \frac{d^4 p'}{(2\p)^4} (2\p) \d^{(+)} (p'^2)        \nonumber \\ 
      \times && \frac{d^4 l}{(2\p)^4} (2\p) \d^{(+)} (l^2)
                \; (2\p)^4 \d^{(4)} (k-p-p')                        \nonumber \\ 
      \times && \big (1-\fm(p_0) \big )\big (1-\fm(p_0') \big)      \nonumber \\ 
      \times && \big (1/2-\fm(l_0) \big )
                \; (\bar \cm\;\cm^*+\bar \cm^* \cm)_{\mbox{\scriptsize (\fref{f:scpes})}}   \; .
\label{eq:scpes}
\eea
Note that the mass-shell constraining delta function of $p$ in these
equations has a derivative because of the double pole due to the 
double propagator $(1/p^2)^2$. The sum over final spins and averaged
over initial spins convoluted amplitudes are
\bea && (\bar \cm\;\cm^*+\bar \cm^* \cm)_{\mbox{\scriptsize (\fref{f:abems})}} 
     \nonumber \\
    = && \frac{16 e^2 g^2}{3} 
       \Big \{ p\cdot p' +(p\cdot q)(p'\cdot q)                    \nonumber \\
      && \mbox{\hspace{2.0cm}} \times 
        \Big (\frac{1}{p^2+2 p\cdot q} +\frac{1}{p^2-2 p\cdot q} \Big ) \Big \} \; ,
\eea 
and 
\bea && (\bar \cm\;\cm^*+\bar \cm^* \cm)_{\mbox{\scriptsize (\fref{f:scpes})}} 
     \nonumber \\
    = && -\frac{8 e^2 g^2}{3} 
       \Big ( 2 (p\cdot l)(p\cdot p') -p^2 (p'\cdot l) \Big )      \nonumber \\
      && \mbox{\hspace{1.0cm}} \times 
       \Big (\frac{1}{2 p\cdot l+p^2} +\frac{1}{2 p\cdot l-p^2} \Big ) \; .       
\eea 

Similar to what was done to \fref{f:2lses2}, one can also put only 
two lepton lines of \fref{f:2lsev1} on mass-shell as shown in 
\fref{f:2lsev2}. In this case, the blob on the r.h.s. of \fref{f:se}
has an internal loop consists of three lines. Converting the photon 
to a real one results in the photon absorption-emission interference
contribution in \fref{f:abemv}. Instead of a forward scattering on
the lepton in \fref{f:abems}, now different fermions participate
in the emission and absorption. Next either lepton line within the
loop can be put on-shell. One gets interference contributions where
the vector particle gets absorbed by a lepton before a dilepton 
pair is radiated off at the end and others where the vector fuses
with a virtual photon radiated from a thermal lepton to form
a dilepton pair. These are shown in \fref{f:scpev}. All these interfere
again with the less interesting simple decay graph with a spectator. 

\bfi
\centerline{
\epsfig{figure=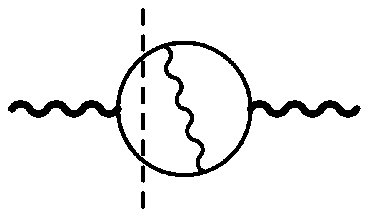,width=3.0cm}}
\caption{Only two lines will be external lines in this two-loop 
self-energy and therefore one of the blobs will have an internal loop 
made up of three lines.} 
\label{f:2lsev2}
\null
\centerline{
\epsfig{figure=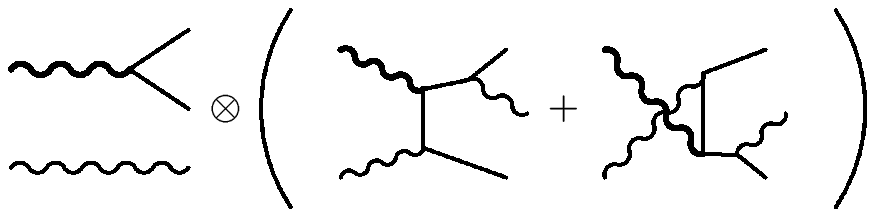,width=7.50cm}}
\caption{From the internal loop of \fref{f:2lsev2}, one also gets a
genuine interference contribution with photon absorption-emission 
within a single graph.} 
\label{f:abemv}
\null
\centerline{
\epsfig{figure=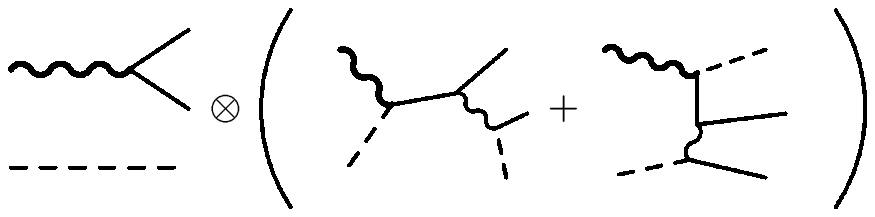,width=7.50cm}}
\caption{A similar real interference contribution to \fref{f:scpes}
from \fref{f:2lsev2}. Note that the former internal but now on-shell 
lepton line is different from that in \fref{f:2lsev2}. There is another
similar contribution from \fref{f:2lsev2} by opening up the other
lepton line in the loop.} 
\label{f:scpev}
\efi

The mathematical expressions for the contributions in 
\ftwref{f:abemv}{f:scpev} are similar to those in
\etwref{eq:abems}{eq:scpes} except {\bf 1)} $(-1)(2\p){\d'}^{(+)}(p^2)$
is replaced by the usual $(2\p)\d^{(+)}(p^2)$ and {\bf 2)}
the convoluted amplitudes are of course not the same. These 
work out to be 
\bea && (\bar \cm\;\cm^*+\bar \cm^* \cm)_{\mbox{\scriptsize (\fref{f:abemv})}} 
                                                               \nonumber \\
    = && \frac{16 e^2 g^2}{3} 
       \Big \{ 1 - \frac{(p\cdot p')^2}{(p\cdot q)(p'\cdot q)} \Big \} \; ,
\eea 
and 
\bea && (\bar \cm\;\cm^*+\bar \cm^* \cm)_{\mbox{\scriptsize (\fref{f:scpev})}} 
     \nonumber \\
    = && -\frac{4 e^2 g^2}{3}\; \frac{p'\cdot l}{p \cdot l}\;       
       \Big \{ \frac{p\cdot l-p\cdot p'}{k\cdot l-p\cdot p'} 
              +\frac{p\cdot l+p\cdot p'}{k\cdot l+p\cdot p'} \Big \} \; . 
\eea 

The full result up to $\co(e^2)$ correction is 
\bea \Gamma &=&   \Gamma_{\mbox{\scriptsize LO}}
              +2\;\Gamma_{\mbox{\scriptsize compton}}
              +   \Gamma_{\mbox{\scriptsize decay}}
              +   \Gamma_{\mbox{\scriptsize fusion}}            \nonumber \\
           & &+2\;\Gamma_{\mbox{\scriptsize (\fref{f:abems})}}
              +2\;\Gamma_{\mbox{\scriptsize (\fref{f:scpes})}}  \nonumber \\
           & &+ \;\Gamma_{\mbox{\scriptsize (\fref{f:abemv})}}  
              +2\;\Gamma_{\mbox{\scriptsize (\fref{f:scpev})}} \; .
\label{eq:gamma}
\eea
The factor of two for the various contributions to the width 
is due to the two possibilities of lepton and antilepton to take part
in whatever interactions that give rise to that particular contribution. 
Note that in the $\Gamma_{\mbox{\scriptsize decay}}$, 
$\Gamma_{\mbox{\scriptsize fusion}}$ and
$\Gamma_{\mbox{\scriptsize \fref{f:abemv}}}$, each has linear infrared
divergence enhanced by the Bose-Einstein distribution but these
cancel in the sum of the three \cite{kw1,wel}. 

By opening up the internal loops, each term in \eref{eq:gamma} is now a 
contribution from a clear physical process and not just a vague one-loop 
and two-loop contribution which cannot be readily associated with an 
interaction. In the general case of self-energy graphs at high orders so that
there are multiple internal loops and overlapping lines, the imaginary-time 
formalism is just too compact for the physics to be transparent. In any case 
if one is concerned with the physical processes, a lot of terms in the 
self-energy will not contribute and drop out once the discontinuity has 
been taken. Therefore there will be a lot of fruitless labor, so instead 
of calculating hoops and loops of virtual particles, it is much 
simpler to disentangle all of them and calculate instead the amplitudes 
of on-shell particle interactions. We have shown here that these could 
be both squared modulus processes and purely interference processes made
possible by the presence of the heat bath. The latters are not so 
well-known and it is easy to be influenced by the accustomed vacuum 
picture to believe that there are only the former contributions. 

Using these techniques we have, beside working out the Z boson width 
in a quark-gluon plasma \cite{kw1}, also calculated the high mass 
next-to-leading order dilepton production from such a QCD plasma. 
This is reported in \cite{kw2}. Some of these have previously 
been examined in for example \cite{aa2}. Since we have not included
any form of resummation, we do not expect the results to agree. 
But there should be an agreement once the resummed version of the
present work is done. We will leave this as a future work. 

As a final remark although we exclusively worked within the 
imaginary-time formalism, the result for any of the physical 
quantities on the LHS in \eforef{eq:pb}{eq:pf}{eq:db}{eq:df} 
obtained within the real-time 
formalism should, of course, be the same. The connection between 
the real- and imaginary-time formalism for the first few N-point 
functions have been worked out by various authors
\cite{ks2,ks1,kob1,kob2,evans,ab,tay}. The relevant quantities
in our case are the imaginary part of the two-point functions. 
These are related to those in the real-time formalism by
\bea  \mbox{Im}\, \Pi    (k) &=&+\tanh (k_0 /2 T)\; 
      \mbox{Im}\, \Pi_{11} (k)    \\
      \mbox{Im}\, \Sigma (k) &=&-\coth (k_0 /2 T)\; 
      \mbox{Im}\, \Sigma_{11} (k)  \; .
\eea
To calculate the imaginary part of the self-energies in real-time, 
one can use any of the many suggested finite temperature cutting rules 
\cite{ks2,wel2,ls,ks1,kob2,bdn,gel}. Since we are more interested in what
physical processes actually contribute to the production or decay rate
of a non-thermalized particle than the mathematical rules by which
one calculates the imaginary part of $\Pi(k)$ or $\Sigma(k)$,
we will leave this at that and do not elaborate any further. 
For the details of the precise connection between any aspects of
the two formalism, the readers should consult the references given
here. The message that we would like to convey in this 
paper is that the physical processes that contribute to the LHS of 
\eforef{eq:pb}{eq:pf}{eq:db}{eq:df} are far richer in numbers and
stranger than those found at zero temperature. The existence
of a thermal bath permits many purely interference processes to
contribute which have no counterpart in the vacuum. It has also
been pointed out that it was possible to calculate the rates,
provided that the finite temperature interference processes were
included, by basing on the actual physical processes rather than on 
which N-point or N-loop functions. The latter are not 
immediately physically transparent.

\section*{Acknowledgments}

The author thanks Joe Kapusta for useful discussions. 
This work was supported by the U.S. Department of Energy under 
grant no. DE-FG02-87ER40328.

\end{document}